\documentclass[onecolumn,10pt]{article}
\usepackage[top=.75in, bottom=.75in, left=.75in, right=.75in]{geometry}
\usepackage{amsmath,amsfonts,amscd,amssymb} 
\usepackage{graphicx}
\usepackage{epstopdf}
\usepackage{overpic}
\usepackage{cancel}
\usepackage{rotating}
\usepackage{url}
\usepackage{caption}
\usepackage{color}
\usepackage{rotating}
\usepackage{multirow}
\usepackage{wrapfig}
\usepackage{mathtools}
\usepackage{subeqnarray}
\usepackage{setspace}
\usepackage{palatino} 
\usepackage[numbers,sort&compress]{natbib}
\usepackage[bottom,flushmargin,hang,multiple]{footmisc}
\usepackage{lipsum}

\definecolor{header1}{cmyk}{0,0,0,1}

\usepackage[utf8]{inputenc}

\usepackage[normalem]{ulem}
\usepackage{color}

\title{\vspace{-.25in}{\huge\selectfont \textbf{Enhancing Computational Fluid Dynamics \\ with Machine Learning}}\vspace{-.1in}}

\author{\normalsize{Ricardo Vinuesa$^{1,2*}$ and Steven L. Brunton$^{3}$}\\
\footnotesize{$^1$ FLOW, Engineering Mechanics, KTH Royal Institute of Technology, Stockholm, Sweden}\\
\footnotesize{$^2$ \textcolor{black}{Swedish e-Science Research Centre (SeRC), Stockholm, Sweden}}\\
\footnotesize{$^3$ Department of Mechanical Engineering, University of Washington, Seattle, WA 98195, United States}\\
\footnotesize{$^*$ Corresponding author: Ricardo Vinuesa (rvinuesa@mech.kth.se) \vspace{-.1in}}
}

\date{}

\begin{document}
\maketitle

\vspace{-.2in}
\begin{abstract}

Machine learning is rapidly becoming a core technology for scientific computing, with numerous opportunities to advance the field of computational fluid dynamics. In this Perspective, we highlight some of the areas of highest potential impact, including to accelerate direct numerical simulations, to improve turbulence closure modeling, and to develop enhanced reduced-order models. We also discuss emerging areas of machine learning that are promising for computational fluid dynamics, as well as some potential limitations that should be taken into account.

\end{abstract}

\section{Introduction}\label{sec:intro}

The field of numerical simulation of fluid flows is generally known as computational fluid dynamics (CFD). Fluid mechanics is an area of great importance, both from a \textcolor{black}{ scientific} perspective and for \textcolor{black}{ a range of industrial-engineering} applications.
Fluid flows are governed by the Navier--Stokes equations, which are partial differential equations (PDEs) modeling the conservation of mass and momentum in a Newtonian fluid. 
These PDEs are \textcolor{black}{ non-linear} due to the convective acceleration (which is related to the change of velocity with the position), and they commonly exhibit time-dependent chaotic behavior, known as {\it turbulence}. 
Solving the Navier--Stokes equations for turbulent flows requires numerical methods that may be computationally expensive, or even intractable at high Reynolds numbers, due to the wide range of scales in space and time necessary to resolve these flows.  There are various approaches to numerically solve these equations, which can be discretized using methods of different orders, for instance finite-difference~\cite{godunov1959finite}, finite-volume~\cite{eymard2000finite}, finite-element~\cite{zienkiewicz1977finite}, spectral methods~\cite{canuto2012spectral}, and so forth. 
Furthermore, turbulence can be simulated with different levels of fidelity and computational cost. 

\textcolor{black}{At the same time, we are experiencing a revolution in the field of machine learning (ML), which is enabling advances across a wide range of scientific and engineering areas~\cite{Brunton2019book,recht2019tour,vinuesa_et_al_2020,Noe2020ARPC,niederer2021scaling}.} \textcolor{black}{ Machine learning is a subfield of the broader area of artificial intelligence (AI), which is focused on the development of algorithms with the capability of learning from data without explicit mathematical models~\cite{samuel}. Many of the most exciting advances in ML have leveraged deep learning, based on neural networks (NNs) with multiple hidden layers between the input and the output. One key aspect contributing to the remarkable success of deep learning is the ability to learn in a hierarchical manner: while initial layers learn simple relationships in the data, deeper layers combine this information to learn more abstract relationships. Many physical problems exhibit this hierarchical behavior, and can therefore be effectively modelled using deep learning, and machine learning more generally.} 

\begin{figure}
\centering 
\includegraphics[width=0.9\textwidth]{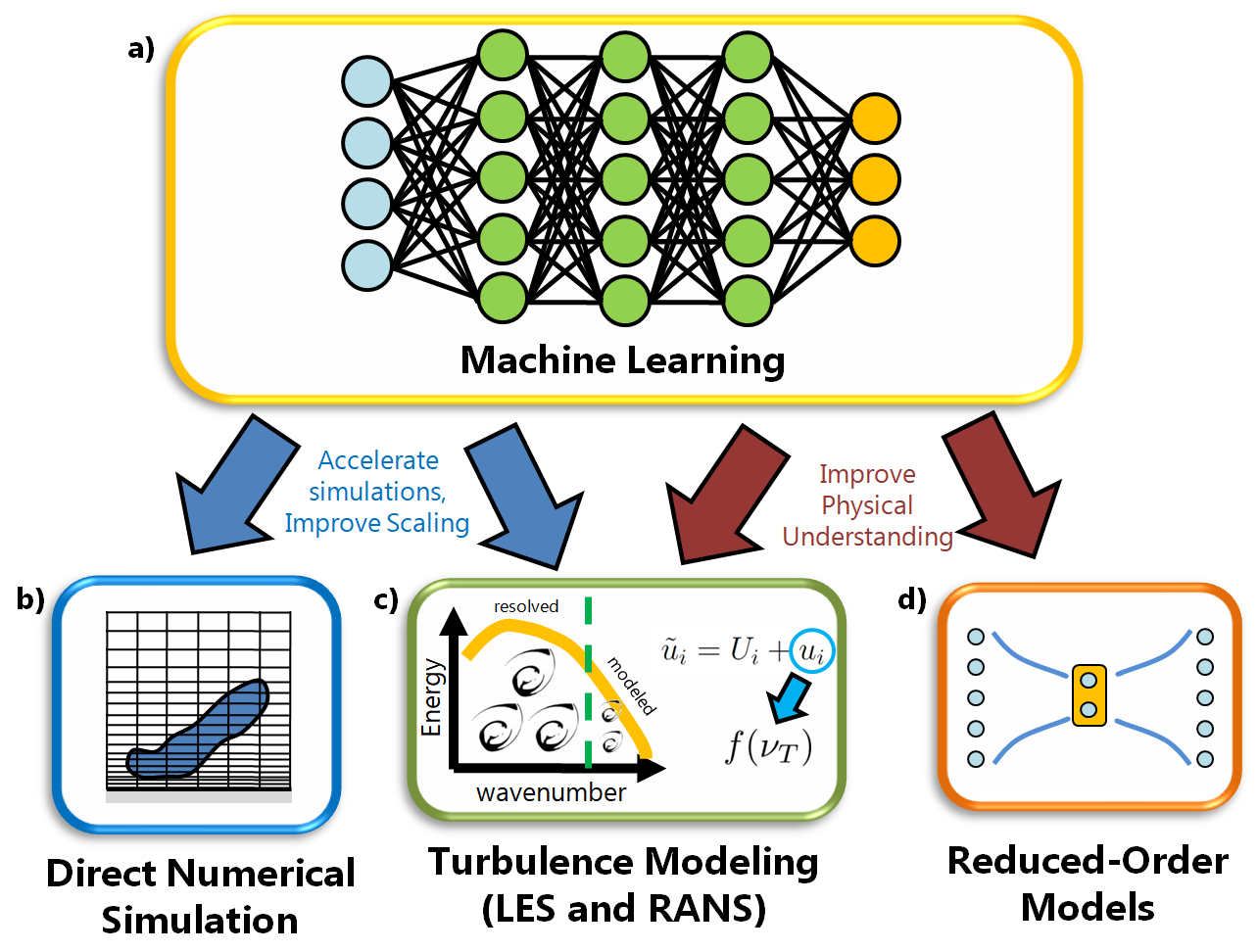}
   \caption{{\bf Summary of some of the most relevant areas where machine learning can enhance CFD.} a) Neural network illustrating the field of machine learning. b) In direct numerical simulations, computational grids fine enough to properly resolve the details of the flow structures (such as the one shown in blue) are needed. The grey band denotes the wall. c) The left subpanel illustrates large-eddy simulation (LES), and we show a schematic of the turbulence spectrum, where the larger scales (to the left of the green vertical line, {\it i.e.} the cut-off wavenumber) are numerically simulated, and the smaller ones (to the right of the cut-off) are represented by a model. Since these smaller scales do not need to be resolved, the computational cost of the simulation is reduced. The right subpanel illustrates Reynolds-averaged Navier--Stokes (RANS) simulations, where $\tilde{u}_i$ is the instantaneous velocity component $i$, $U_i$ the mean and $u_i$ the fluctuating component. In RANS models, the fluctuations are usually expressed as a function of the eddy viscosity $\nu_T$. d) General schematic of a dimensionality-reduction method, where the circles denote flow quantities. The input (on the left in blue) is the high-resolution flow field, the output (on the right in red) is the reconstruction of the input field, and the yellow box represents the low-dimensional system.}
   \label{fig:fig1}
\end{figure}

In this Perspective, we focus on the potential of machine learning to improve CFD, including possibilities to increase the speed of high-fidelity simulations, develop turbulence models with different levels of fidelity, and produce reduced-order models beyond what can be achieved with classical approaches. 
Several authors have surveyed the potential of machine learning to improve fluid mechanics ~\cite{Brenner2019prf,Brunton2020arfm}, including topics beyond the scope of CFD, such as experimental techniques, control applications, and related fields. 
Others have reviewed more specific aspects of ML for CFD, such as turbulence closure~\cite{duraisamy_et_al,ahmed2021closures} and heat-transfer aspects of CFD for aerodynamic optimization~\cite{wang_wang}.
Our discussion will address the middle ground of ML for CFD more broadly, with a schematic representation of topics covered in Fig.~\ref{fig:fig1}.
Approaches to improve CFD with ML are aligned with the larger efforts to incorporate ML into scientific computing, for instance via physics-informed neural networks (PINNs)~\cite{raissi2019physics,karniadakis2021physics} or to accelerate computational chemistry~\cite{Noe2019science,Noe2020ARPC}.

\section{Accelerating direct numerical simulations}\label{sec:performance}

Direct numerical simulation (DNS) is a high-fidelity approach where the governing Navier--Stokes equations are discretized and integrated in time with enough degrees of freedom to resolve all flow structures. 
Turbulent flows exhibit a pronounced multi-scale character, with vortical structures across a range of sizes and energetic content~\cite{vinuesa_ftac}. 
This complexity requires fine meshes and accurate computational methods to avoid distorting the underlying physics with numerical artifacts. 
With a properly designed DNS, it is possible to obtain a representation of the flow field with the highest level of detail among CFD methods. 
However, the fine computational meshes required to resolve the smallest scales lead to exceedingly high computational costs, which increase with the Reynolds number~\cite{choi_moin}. 

\begin{figure}[t]
\centering 
\includegraphics[width=0.95\textwidth]{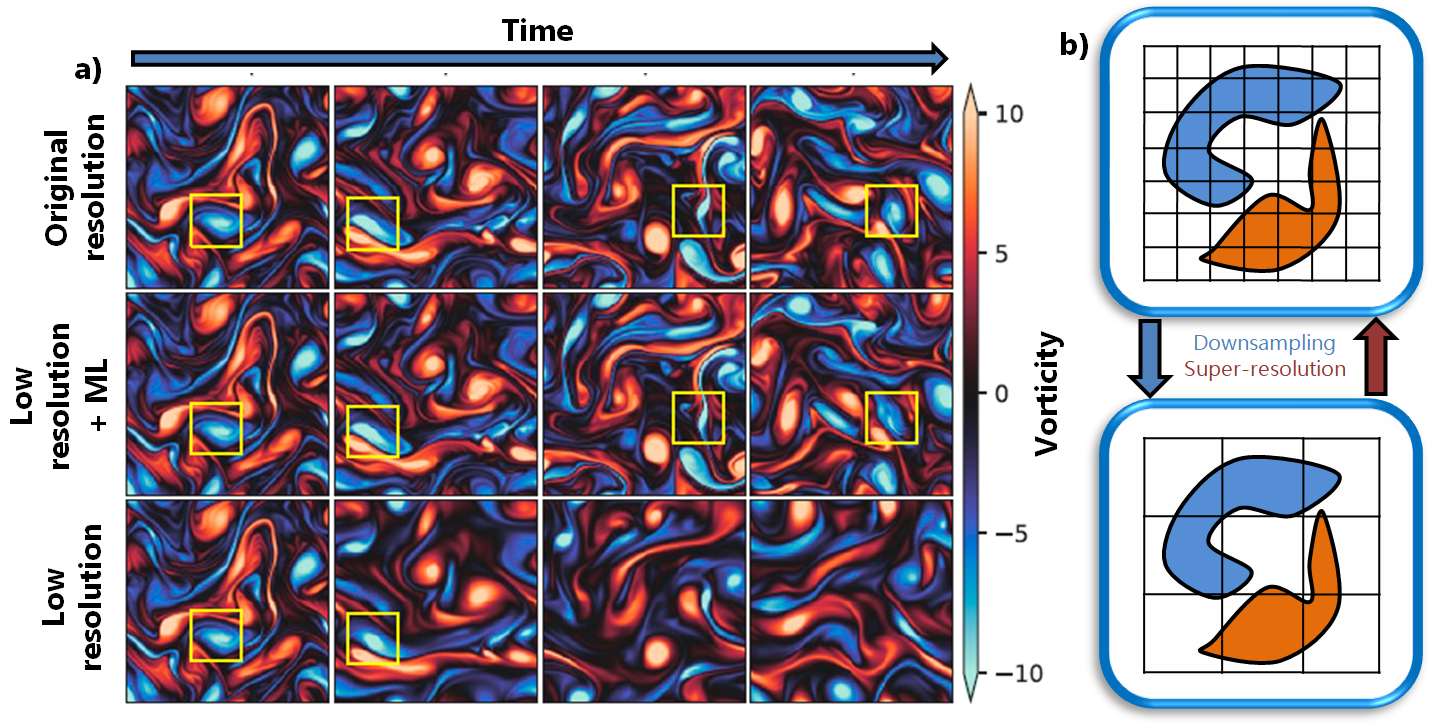}
\vspace{-.1in}
   \caption{{\bf An example of machine-learning-accelerated direct numerical simulation.} a) Results from the work by Kochkov~{\it et al.}~\cite{hoyer}, where the instantaneous vorticity field is shown for simulations with high-/low-resolution, as well as a case with low resolution supplemented with ML. Four different time steps are shown, and some key vortical structures are highlighted with yellow squares. Adapted from the work by Kochkov~{\it et~al.}~\cite{hoyer}, with permission of the publisher (United States National Academy of Sciences). b) Sketch represents that ML accelerates the simulation through a correction between the coarse and the fine resolutions. In this example, one can reduce the resolution by downsampling the flow data on a coarser mesh, and then use ML to recover the details present in the finer simulation through super-resolution. The correction between coarse and fine resolutions, which is based on ML, enables the super-resolution task here.}
   \label{fig:fig2}
\end{figure}

A number of machine-learning approaches have been developed recently to improve the efficiency of DNS. First, we will discuss several studies aimed at improving discretization schemes. Bar-Sinai {\it et al.}~\cite{bar2019learning} proposed a technique based on deep learning to estimate spatial derivatives in low-resolution grids, outperforming standard finite-difference methods. A similar approach was developed by Stevens and Colonius~\cite{stevens2020enhancement} to improve the results of fifth-order finite-difference schemes in the context of shock-capturing simulations. Jeon~and~Kim~\cite{jeon_kim} proposed to use a deep neural network to simulate the well-known finite-volume discretization scheme~\cite{eymard2000finite} employed in fluid simulations. 
They tested their method with reactive flows, obtaining \textcolor{black}{very good} agreement with  reference high-resolution data at one tenth the computational cost. However, they also documented errors with respect to the reference solution which increased with time. Another deep-learning approach, based on a fully-convolutional/long-short-term-memory (LSTM) network, was proposed by Stevens~and~Colonius~\cite{stevens2020finitenet} to improve the accuracy of finite-difference/finite-volume methods. Second, we consider the strategy of developing a correction between fine- and coarse-resolution simulations. This was developed by Kochkov~{\it et al.}~\cite{hoyer} for the two-dimensional Kolmogorov flow~\cite{chandler}, which maintains fluctuations via a forcing term. They leveraged deep learning to develop the correction, obtaining excellent agreement with reference simulations in meshes from 8 to 10 times coarser in each dimension, as shown in Fig.~\ref{fig:fig2}. \textcolor{black}{In particular, this figure shows that for long simulations, the standard coarse-resolution case does not exhibit certain important vortical motions, which are however properly captured by the low-resolution case with the ML model.} These results promise to \textcolor{black}{substantially} reduce the computational cost of relevant fluid simulations, including weather~\cite{bauer_et_al}, climate~\cite{freddy}, engineering~\cite{vinuesa_et_al_2018} and astrophysics~\cite{toras2018towards}. Finally, other strategies to improve the performance of PDE solvers in coarser meshes have been developed by Li {\it et al.}~\cite{li2020fourier,li2020multipole,li2020neural}. 

It is also possible to accelerate CFD by solving the Poisson equation with deep learning; this has been done by several research groups in various areas, such as simulations of electric fields~\cite{poisson2} and particles~\cite{poisson}. The Poisson equation is frequently used in operator-splitting methods to discretize the Navier--Stokes equations~\cite{bridson}, where first the velocity field is advected, and the resulting field $\mathbf{u}^*$ does not satisfy the continuity equation ({\it i.e.} for incompressible flows, $\mathbf{u}^*$ does not satisfy the \textcolor{black}{divergence-free} condition). The second step involves a correction to ensure that $\mathbf{u}^*$ is divergence free, leading to the following Poisson equation: 
\begin{equation} \label{ec:poisson}
\frac{\Delta t}{\rho} \nabla^2 p = - \nabla \cdot \mathbf{u}^*,
\end{equation}
where $\Delta t$ is the simulation time step, $\rho$ is the fluid density, and $p$ is the pressure. 
Solving this equation is typically the most computationally expensive step of the numerical solver.  
Therefore, devising alternative strategies to solve it more efficiently is an area of great promise. \textcolor{black}{Machine learning can be used for this task, because this technology can exploit data from previous examples to find mappings between the divergence of the uncorrected velocity and the resulting pressure field. For instance,}
Ajuria~{\it et al.}~\cite{poisson3} proposed using a convolutional neural network (CNN) to solve Equation~(\ref{ec:poisson}) in incompressible simulations, and tested it in a plume configuration. Their results indicate that it is possible to outperform the traditional Jacobi solver with good accuracy at low Richardson numbers $Ri$ (the Richardson number measures the ratio between the buoyancy and the shear in the flow). However, the accuracy degrades at higher $Ri$, motivating the authors to combine the CNN with the Jacobi solver (which is used when the CNN diverges). CNNs were also used to solve the Poisson problem by decomposing the original problem into a homogeneous Poisson problem plus four inhomogeneous Laplace subproblems~\cite{poisson_imperial}. This decomposition resulted in errors below $10\%$, which motivates using this approach as a first guess in iterative algorithms, potentially reducing computational cost. \textcolor{black}{Other data-driven methods~\cite{Weymouth} have been proposed to accelerate the pressure correction in multi-grid solvers.} These approaches may also be used to accelerate simulations of lower fidelity that rely on turbulence models.

Numerical simulations can also be accelerated by decreasing the size of the computational domain needed to retain physical properties of the system. \textcolor{black}{Two ways in which the domain can be reduced are to replace a section upstream of the domain of interest by an inflow condition, and to replace part of the far-field region by a suitable boundary condition. Doing so, these parts of the domain do not need to be simulated, thus producing computational savings, and ML can help to develop the inflow and far-field conditions as discussed next.} Fukami~{\it et al.}~\cite{fukami_inflow} developed a time-dependent inflow generator for wall-bounded turbulence simulations using a convolutional autoencoder with a multilayer perceptron (MLP). 
They tested their method in a turbulent channel flow at $Re_{\tau}=180$, which is the friction Reynolds number based on channel half height and friction velocity, and they maintained turbulence for an interval long enough to obtain converged turbulence statistics. This is a promising research direction due to the fact that current inflow-generation methods show limitations in terms of generality of the inflow conditions, for instance at various flow geometries and Reynolds numbers. A second approach to reduce the computational domain in external flows is to devise a strategy to set the right pressure-gradient distribution without having to simulate the far field. This was addressed by Morita~{\it et al.}~\cite{morita_et_al} through Bayesian optimization based on Gaussian-process regression, achieving accurate results when imposing concrete pressure-gradient conditions on a turbulent boundary layer.

\section{Improving turbulence models}\label{sec:rans}

DNS is impractical for many real-world applications due to the computational cost associated with resolving all scales for flows with high Reynolds numbers, together with difficulties arising from complex geometries. 
Industrial CFD typically relies on either Reynolds-averaged Navier--Stokes (RANS) models, where no turbulent scales are simulated, or coarsely-resolved large-eddy simulations (LES), where only the largest turbulent scales are resolved and smaller ones are modelled. 
Here the term {\it model} refers to an {\it a-priori} assumption regarding the physics of a certain range of turbulent scales. In the following we discuss ML applications to RANS and LES modeling. 

\subsection{RANS modeling}

Numerical simulations based on RANS models rely on the so-called RANS equations, which are obtained after decomposing the instantaneous flow quantities into a mean and a fluctuating component, and averaging the Navier--Stokes equations in time. \textcolor{black}{Using index notation, the instantaneous $i$-th velocity component $\tilde{u}_i$ can be decomposed into its mean ($U_i$) and fluctuating ($u_i$) components as follows: $\tilde{u}_i=U_i+u_i$.} Although the RANS equations govern the mean flow, the velocity fluctuations are also present in the form of the Reynolds stresses $\overline{u_i u_j}$ \textcolor{black}{(where the overbar denotes averaging in time), which are correlations between the $i$-th and $j$-th velocity components.} Since it is convenient to derive equations containing only mean quantities, $\overline{u_i u_j}$ needs to be expressed as a function of the mean flow, and this is called the {\it closure problem}. The first approach to do this was proposed by Boussinesq~\cite{boussinesq}, who related the Reynods stresses with the mean flow via the so-called eddy viscosity $\nu_T$. Although this approach has led to some success, there are still a number of open challenges for RANS modeling in complex flows~\cite{NASA}, where this approach is too simple. \textcolor{black}{In particular, conventional RANS models exhibit significant errors when dealing with complex pressure-gradient distributions, complicated geometries involving curvature, separated flows, flows with significant degree of anisotropy and three-dimensional effects, among others. As argued by Kutz~\cite{kutz2017deep}, machine learning can produce more sophisticated models for the Reynolds stresses by using adequate data, in particular if the examples used for training represent a sufficiently rich set of flow configurations.}

A wide range of ML methods have recently been used to improve RANS turbulence modeling~\cite{duraisamy_et_al}, \textcolor{black}{focusing on the challenge of improving the accuracy of the Reynolds stresses for general conditions. For example, Ling~{\it et al.}~\cite{ling2016reynolds} proposed} a novel architecture, including a multiplicative layer with an invariant tensor basis, used to embed Galilean invariance in the predicted anisotropy tensor. 
Incorporating this invariance improves the performance of the network, which outperforms traditional RANS models based on linear~\cite{boussinesq} and nonlinear~\cite{craft_et_al} eddy-viscosity models.  
They tested their models for turbulent duct flow and the flow over a wavy wall, which are challenging to predict with RANS models~\cite{hexagon,spalart} because of the presence of secondary flows~\cite{wavy}. 
Other ML-based approaches~\cite{wang2017physics,wu_et_al} rely on physics-informed random forests to improve RANS models, with applications to cases with secondary flows and separation. \textcolor{black}{On the other hand, Jiang~{\it et al.}~\cite{ml_rans} recently developed an interpretable framework for RANS modeling based on a physics-informed residual network (PiResNet).} Their approach relies on two modules to infer the structural and parametric representations of turbulence physics, and includes non-unique mappings, a realizability limiter, and noise-insensitivity constraints. 
\textcolor{black}{Interpretable models are essential for engineering and physics, particularly in the context of turbulence modeling. The interpretability of the framework by Jiang~{\it et al.}~\cite{ml_rans} relies on its constrained model form~\cite{rudin}, although this is not generally possible to attain. Recent studies~\cite{vinuesa_interp} have discussed various approaches to include interpretability in the development of deep-learning models, and one promising approach is the one proposed by Cranmer {\it et al.}~\cite{cranmer_et_al}. This technique has potential in terms of embedding physical laws and improving our understanding of such phenomena.} Other interpretable RANS models were proposed by Weatheritt~and~Sandberg~\cite{sandberg1}, using gene-expression programming (GEP), which is a branch of evolutionary computing~\cite{gep}. 
GEP iteratively improves a population of candidate solutions by survival of the fittest, with the advantage of producing closed-form models. The Reynolds-stress anisotropy tensor was also modelled by Weatheritt~and~Sandberg~\cite{sandberg2}, who performed tests in RANS simulations of turbulent ducts. Models based on sparse identification of nonlinear dynamics (SINDy)~\cite{Brunton2016pnas} have also been used for RANS closure models~\cite{beetham2020formulating,schmelzer2020discovery,beetham2021sparse}. 

\textcolor{black}{Furthermore, the} literature also reflects the importance of imposing physical constraints in the models and incorporating uncertainty-quantification (UQ)~\cite{rezaeiravesh,iaccarino1,iaccarino2} alongside ML-based models. It is also important to note that when using DNS quantities to replace terms in the RANS closure, the predictions may be unsatisfactory~\cite{poroseva_et_al}. 
This inadequacy is due to the strict assumptions associated with the RANS model, as well as the potential ill-conditioning of the RANS equations~\cite{wu_et_al_2018}. 
It is thus essential to take advantage of novel data-driven methods, while also ensuring that uncertainties are identified and quantified. 
Another interesting review by Ahmed~{\it et al.}~\cite{ahmed2021closures} discussed both classical and emerging data-driven closure approaches, also connecting with ROMs.

Obiols-Sales~{\it et al.}~\cite{accelerator} developed a method to accelerate the convergence of RANS simulations based on the very popular Spalart--Allmaras (SA) turbulence model~\cite{sa} using the CFD code OpenFOAM~\cite{openfoam}. 
In essence, they combined iterations from the CFD solver and evaluation of a CNN model, obtaining convergence from 1.9 to 7.4 times faster than that of the CFD solver, both in laminar and turbulent flows. 
Multiphase flows, which consist of flows with two or more thermodynamic phases, are also industrially relevant.  
Gibou~{\it et al.}~\cite{multiphase} proposed different directions in which machine learning and deep learning can improve CFD of multiphase flows, in particular when it comes to enhancing the simulation speed. 
Ma~{\it et al.}~\cite{ma_et_al} used deep learning to predict the closure terms (i.e., gas flux and streaming stresses) in their two-fluid bubble flow, whereas Mi~{\it et al.}~\cite{mi_et_al} analyzed gas-liquid flows and employed neural networks to identify the different flow regimes.

\subsection{LES modeling}

LES-based numerical simulations rely on low-pass filtering the Navier--Stokes equations (Fig.~\ref{fig:fig1}c), such that the largest scales are simulated and the smallest ones (below a certain cut-off) are modeled by means of a so-called subgrid-scale model (SGS). Note that the smallest scales are the most demanding from a computational point of view \textcolor{black}{(both in terms of computer time and memory usage),} because they require fine meshes to be properly resolved. The first proponent of this type of approach was Smagorinski~\cite{smagorinski}, who developed an SGS model based on an eddy viscosity which was computed in terms of the mean flow and the grid size; his model assumed that the production equals the dissipation for the small scales. Although LES can lead to \textcolor{black}{substantial} computational savings with respect to DNS while exhibiting a good accuracy, \textcolor{black}{there are still challenges associated with its usage for general purpose~\cite{NASA}. For instance, current SGS models exhibit limited accuracy in their predictions of turbulent flows at high Reynolds number, in complex geometries, and in cases with shocks and chemical reactions.}

\begin{figure}[t]
\centering 
\includegraphics[width=0.975\textwidth]{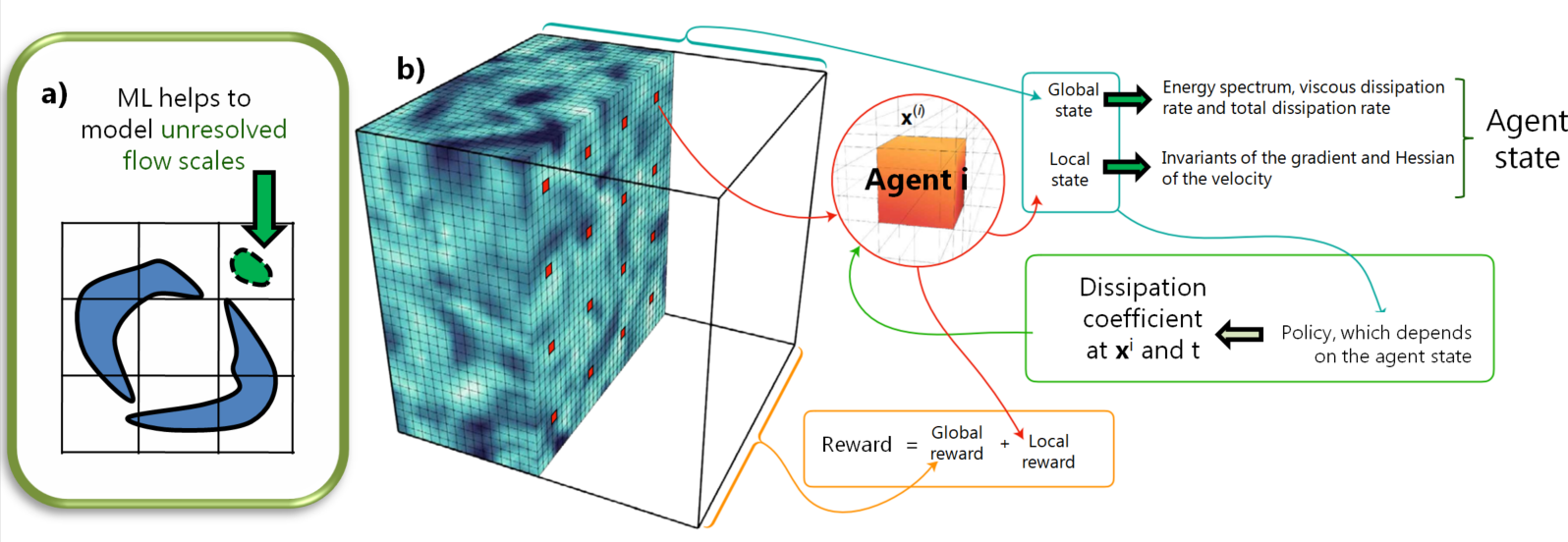}
   \caption{{\bf An example of LES modeling where the dissipation coefficient in the Smagorinski model is calculated by means of machine learning.} a) Schematic of a numerical simulation fine enough to resolve the blue flow structures, but too coarse to resolve the green structure. This small-scale structure (green) would have to be modeled by means of a subgrid-scale (SGS) model, such as the well-known SGS model by Smagorinski~\cite{smagorinski}. This model relies on an empirical parameter, the dissipation coefficient $C_s$,  which can be determined more accurately by ML. b) An example of ML-based approach to determine the value of $C_s$ by reinforcement learning (RL)~\cite{petros_natmi}. The RL agents are located at the red blocks, and they are used to compute $C_s$ for each grid point with coordinates ${\bf x}^i$ at time $t$. This calculation is carried out in terms of a policy, which depends on the state of the agent. This state is based on local variables (invariants of the gradient and Hessian of the velocity) and also global ones (energy spectrum, viscous dissipation rate and total dissipation rate). The agents receive a reward, which also depends on local and global quantities, based on the accuracy of the simulation. Adapted from the work by Novati~{\it et~al.}~\cite{petros_natmi}, with permission of the publisher (Springer Nature).}
   \label{fig:fig4}
\end{figure}

Machine learning has also been used to develop SGS models in the context of LES of turbulent flows \textcolor{black}{basically in two ways: supplementing the unresolved energy in the coarse mesh using supervised learning and developing agent-based approaches to stabilize the coarse simulation. When it comes to the first approach, we list several studies below which rely on high-fidelity data to train both deep-learning and GEP-based models. First,} Beck~{\it et al.}~\cite{beck_et_al} used an artificial neural network based on local convolutional filters to predict the mapping between the flow in a coarse simulation and the closure terms, using a filtered DNS of decaying homogeneous isotropic turbulence. 
Lapeyre~{\it et al.}~\cite{lapeyre_et_al} employed a similar approach, with a CNN architecture inspired by a U-net model, to predict the subgrid-scale wrinkling of the flame surface in premixed turbulent combustion; they obtained better results than classical algebraic models. 
Maulik~{\it et al.}~\cite{maulik2019subgrid} employed a multilayer perceptron (MLP) to predict the SGS model in an LES using high-fidelity numerical data to train the model. 
They evaluated the performance of their method on Kraichnan turbulence~\cite{kraichnan}, which is a classical two-dimensional decaying-turbulence test case. Several other studies have also used neural networks in a supervised manner for SGS modeling~\cite{vollant,gamahara,maulik}. Furthermore, GEP has also been used for SGS modeling~\cite{sandberg3} in an LES of a Taylor--Green vortex, outperforming standard LES models. \textcolor{black}{Regarding the second approach to LES modeling,} Novati~{\it et al.}~\cite{petros_natmi} employed multi-agent reinforcement-learning (RL) to estimate the unresolved subgrid-scale physics. 
This unsupervised method exhibits favorable generalization properties across grid sizes and flow conditions, and the results are presented for isotropic turbulence. \textcolor{black}{As shown in the schematic representation of Fig.~\ref{fig:fig4}, the state of the agents at a particular instant is given in terms of both local and global variables; this state is then used to calculate the so-called dissipation coefficient at each grid point.} 

In certain applications, for instance those involving atmospheric boundary layers (ABLs), the Reynolds number is several orders of magnitude larger than those of most studies based on turbulence models or wind-tunnel experiments~\cite{hutchins}. 
The mean flow in the inertial sublayer has been widely studied in the ABL community, and it is known that in neutral conditions it can be described by a logarithmic law~\cite{log}. 
The logarithmic description of the inertial sublayer led to the use of wall models, which replace the region very close to the wall with a model defining a surface shear stress matching the logarithmic behavior. 
This is the cornerstone of most atmospheric models, which avoid resolving the computationally expensive scales close to the wall. 
One example is the work by Giometto~{\it et al.}~\cite{giometto}, who studied a real urban geometry, adopting the LES model by Bou-Zeid~{\it et al.}~\cite{bou_zeid} and the Moeng model~\cite{moeng} for the wall boundary condition. \textcolor{black}{It is possible to use data-driven approaches to develop mappings between the information in the outer region (which is resolved) and a suitable off-wall boundary condition (so the near-wall region does not need to be resolved).} For instance, it is possible to exploit properties of the logarithmic layer and rescale the flow in the outer region to set the off-wall boundary condition in turbulent channels~\cite{mizuno_jimenez,gmayoral}. 
This may also be accomplished via transfer functions in spectral space~\cite{sasaki}, convolutional neural networks~\cite{ari_etmm}, or modeling the temporal dynamics of the near-wall region via deep neural networks~\cite{milano_koumoutsakos}. Other promising approaches based on deep learning are the one by Moriya~{\it et al.}~\cite{moriya}, based on defining a virtual velocity, and the reinforcement-learning technique by Bae~and~Koumoutsakos~\cite{bae_koumoutsakos}. Defining off-wall boundary conditions with machine learning is a challenging yet promising area of research.

\section{Developing reduced-order models}\label{sec:roms}
Machine learning is also being used to develop reduced-order models (ROMs) in fluid dynamics. 
ROMs rely on the fact that even complex flows often exhibit a few dominant coherent structures~\cite{Taira2017aiaa,Rowley2017arfm,Taira2020aiaaj} that may provide coarse, but valuable information about the flow. 
Thus, ROMs describe the evolution of these coherent structures, providing a lower-dimensional, lower-fidelity characterization of the fluid. 
In this way, ROMs provide a fast surrogate model for the more expensive CFD techniques described above, enabling optimization and control tasks that rely on many model iterations or fast model predictions. 
The cost of this efficiency is a loss of generality: ROMs are tailored to a specific flow configuration, providing massive acceleration but a limited range of applicability.  

Developing a reduced-order model involves (1) finding a set of reduced \emph{coordinates}, typically describing the amplitudes of important flow structures, and (2) identifying a differential-equation model ({\it i.e.}, a dynamical system) for how these amplitudes evolve in time. 
Both of these stages have seen incredible recent advances with machine learning.  
One common ROM technique involves learning a low-dimensional coordinate system with the proper orthogonal decomposition (POD)~\cite{lumley,Taira2017aiaa} and then obtaining a dynamical system for the flow system restricted to this subspace by Galerkin projection of the Navier--Stokes equations onto these modes.  
Although the POD step is data driven, working equally well for experiments and simulations, Galerkin projection requires a working numerical implementation of the governing equations; moreover, it is often \emph{intrusive}, involving custom modifications to the numerical solver.  
The related dynamic-mode decomposition (DMD)~\cite{Schmid2010jfm} is a purely data-driven procedure that identifies a low-dimensional subspace and a linear model for how the flow evolves in this subspace. Here, we will review a number of recent developments to extend these approaches with machine learning.  

\begin{figure}[t]
    \centering
        \includegraphics[width=\textwidth]{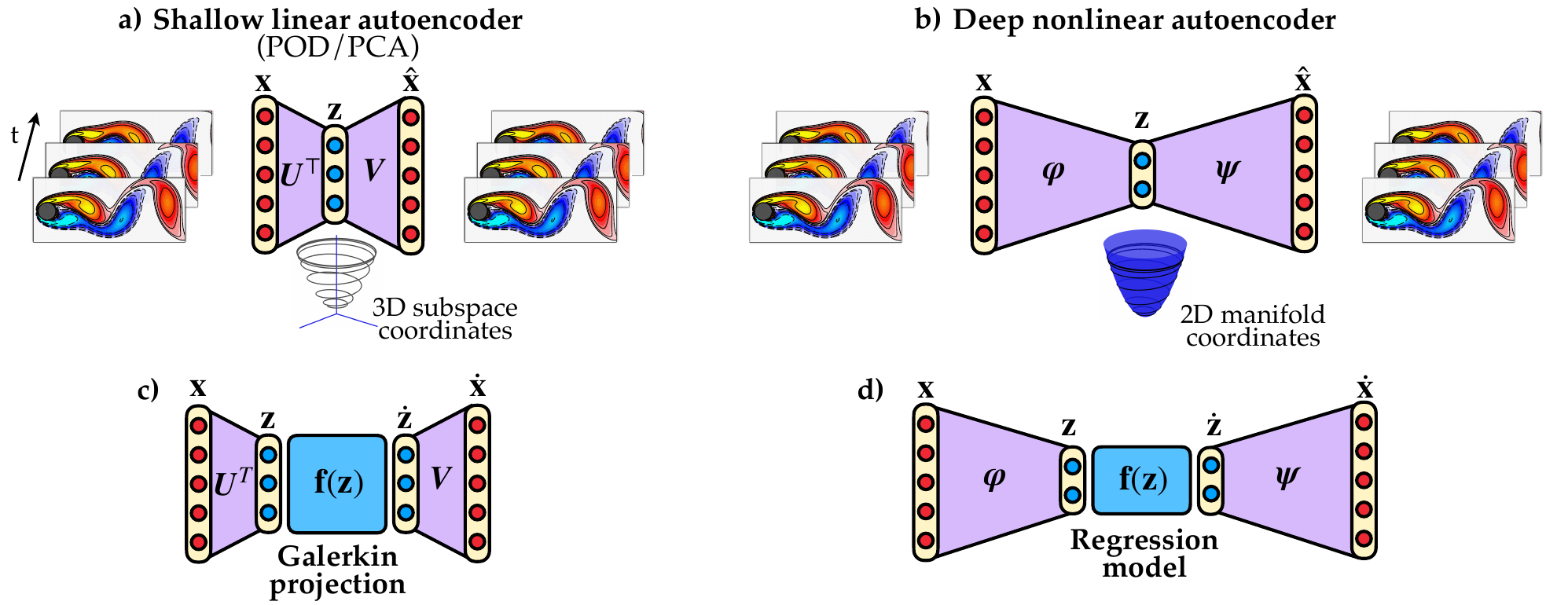}
    \caption{{\bf Schematic of neural-network autoencoders for dimensionality reduction and model identification.} Reduced-order models are typically designed to balance efficiency and accuracy. ML solutions further improve the efficiency by reducing the effective dimension of the model and increasing the accuracy through better modeling of how these few variables co-evolve in time.
    In this figure, the input is a high-resolution flow field evolving in time ($t$) and the output is a reconstruction of that field from the latent space. a) Classic proper orthogonal decomposition/principal component analysis (POD/PCA) may be viewed as a shallow autoencoder with a single encoder $\mathbf{U}^\top$ and decoder $\mathbf{V}$ layers, together with linear activation units. For the flow past a cylinder example shown, the dynamics evolve in a three-dimensional coordinate system.  b) A deep, multi-level autoencoder with mulit-layer encoder $\boldsymbol{\varphi}$ and decoder $\boldsymbol{\psi}$, as well as nonlinear activation functions provides enhanced nonlinear coordinates on a manifold.  The cylinder flow now evolves on a two-dimensional submanifold. c) The classic Galerkin-projection model, obtained by projecting the governing Navier--Stokes equations onto an orthogonal basis.  The Galerkin-projection model in c) can be replaced by more generic machine-learning regressions in d), such as long-short-term-memory (LSTM) networks, reservoir networks or sparse nonlinear models to represent the nonlinear dynamical system $\dot{\mathbf{z}} = \mathbf{f}(\mathbf{z})$.}
    \label{fig:ROMfig}
\end{figure}

The first broad opportunity to incorporate machine learning into ROMs is in learning an improved coordinate system in which to represent the reduced dynamics.  
POD~\cite{lumley,Taira2017aiaa} provides an orthogonal set of modes that may be thought of as a data-driven generalization of Fourier modes, which are tailored to a specific problem. 
POD is closely related to principal-component analysis (PCA) and the singular-value decomposition (SVD)~\cite{Brunton2019book}, which are two core dimensionality-reduction techniques used in data-driven modeling. 
These approaches provide linear subspaces to approximate data, even though it is known that many systems evolve on a nonlinear manifold. 
Deep learning provides a powerful approach to generalize the POD/PCA/SVD dimensionality reduction from learning a linear subspace to learning coordinates on a curved manifold. 
Specifically, these coordinates may be learned using a neural network  \emph{autoencoder}, which has an input and output the size of the high-dimensional fluid state and a constriction or bottleneck in the middle that reduces to a low-dimensional latent variable.  
The map from the high-dimensional state $\mathbf{x}$ to the latent state $\mathbf{z}$ is called the \emph{encoder} and the map back from the latent state to an estimate of the high-dimensional state $\hat{\mathbf{x}}$ is the \emph{decoder}. 
The autoencoder loss function is $\|\hat{\mathbf{x}}-\mathbf{x}\|_2^2$.  
When the encoder and decoder consist of a single layer and all nodes have identity activation functions, then the optimal solution to this network will be closely related to POD~\cite{baldi1989neural}. However, this shallow linear autoencoder may be generalized to a deep nonlinear autoencoder with multiple encoding and decoding layers and nonlinear activation functions for the nodes. In this way, a deep autoencoder learns nonlinear manifold coordinates that may considerably improve the compression in the latent space, with increasing applications in fluid mechanics~\cite{murata2020nonlinear,ae_modal}.  
This concept is illustrated in Fig.~\ref{fig:ROMfig} for the simple flow past a cylinder, where it is known that the energetic coherent structures evolve on a parabolic sub-manifold in the POD subspace~\cite{Noack2003jfm}.  
Lee and Carlberg~\cite{lee2020model} recently showed that deep convolutional autoencoders may be used to greatly improve the performance of classical ROM techniques based on linear subspaces~\cite{Benner2015siamreview,Carlberg2017jcp}. More recently, Cenedese~{\it et~al.}~\cite{cenedese_et_al} proposed a promising data-driven method to construct ROMs on spectral submanifolds (SSMs).

Once an appropriate coordinate system is established, there are many machine-learning approaches to model the dynamics in these coordinates. Several neural networks are capable of learning \textcolor{black}{nonlinear dynamics~\cite{sole_conv}, including the LSTM network~\cite{vlachas_et_al,srinivasan2019predictions,sole_lstm}} and echo-state networks~\cite{pathak2018model}, which are a form of reservoir computing. 
Beyond neural networks, there are alternative regression techniques to learn effective dynamical-systems models.  
Cluster reduced-order modeling (CROM)~\cite{Kaiser2014jfm} is a simple and powerful unsupervised-learning approach that decomposes a time series into a few representative clusters, and then models the transition-state probability between clusters.  
The operator-inference approach is closely related to Galerkin projection, where a neighboring operator is learned from data~\cite{peherstorfer2016data,benner2020operator,qian2020lift}.  
The sparse identification of nonlinear dynamics (SINDy)~\cite{Brunton2016pnas} procedure learns a minimalistic model by fitting the observed dynamics to the fewest terms in a library of candidate functions that might describe the dynamics, resulting in models that are interpretable and balance accuracy and efficiency.  
SINDy has been used to model a range of fluids~\cite{Loiseau2017jfm,loiseau2020data,guan2020sparse,Deng2020JFM,deng2021galerkin,callaham2021empirical,callaham2021role}, including laminar and turbulent wake flows, convective flows, and shear flows.  

The modeling approaches above may be combined with a deep autoencoder to uncover a low-dimensional latent space, as in the SINDy-autoencoder~\cite{Champion2019pnas}.  
DMD has also been extended to nonlinear coordinate embeddings through the use of deep-autoencoder networks~\cite{Yeung2017arxiv,Takeishi2017nips,Lusch2018natcomm,Mardt2018natcomm,Otto2019siads}. In all of these architectures, there is a tremendous opportunity to embed partial knowledge of the physics, such as conservation laws~\cite{Loiseau2017jfm}, symmetries~\cite{guan2020sparse}, and invariances~\cite{wang2020incorporating,wang2020towards,frezat2021physical}. 
It may also be possible to directly impose stability in the learning pipeline~\cite{Loiseau2017jfm,erichson2019physics,kaptanoglu2021promoting}. 

The ultimate goal for machine-learning ROMs is to develop models that have improved accuracy and efficiency, better generalizability to new initial and boundary conditions, flow configurations, varying parameters, as well as improved model interpretability, ideally with less intrusive methods and less data.  
Enforcing partially-known physics, such as symmetries and other invariances, along with sparsity, is expected to be critical in these efforts. It is also important to continue integrating these efforts with the downstream applications of control and optimization. Finally, many applications of fluid dynamics involve safety-critical systems, and therefore certifiable models are essential. 

\section{Emerging possibilities and outlook}\label{sec:out}

In this Perspective we have provided our view on the potential of ML to advance the capabilities of CFD,
focusing on three main areas: accelerating simulations, enhancing turbulence models, and improving reduced-order
models. In Table~\ref{table:summary} we highlight some examples of application within each of the three areas. There are also several emerging areas of ML that are promising for CFD, which we discuss in this section. One area is non-intrusive sensing - that is the possibility of performing flow predictions based on, for instance, information at the wall. This task, which has important implications for closed-loop flow control~\cite{drl_control}, has been carried out via CNNs in turbulent channels~\cite{guastoni2}. In connection to the work by Guastoni~{\it et~al.}~\cite{guastoni2}, there are a number of studies documenting the possibility of performing super-resolution predictions ({\it e.g.} when limited flow information is available) in wall-bounded turbulence using CNNs, autoencoders and generative-adversarial networks (GANs)~\cite{kim910unsupervised,fukami2019super,guemes_gans,fukami_autoencoders}. Another promising direction is the imposition of constraints based on physical invariances and symmetries on the ML model, which has been used for SGS modeling~\cite{wang2020incorporating}, ROMs~\cite{Loiseau2017jfm}, and for geophysical flows~\cite{frezat2021physical}.

\begin{table}
  \begin{center}
\def~{\hphantom{0}}
  \begin{tabular}{cccc}
{\bf Accelerate DNS} & {\bf Improve LES/RANS} & {\bf Develop ROMs} & {\bf Future developments} \\
\hline
Deep learning for correction & Deep learning /  & Proper-orthogonal   & Deep learning for  \\
between coarse and & reinforcement learning & decomposition   & prediction and control   \\
fine simulations & for more general & for improved linear subspaces  & of turbulent
flows
 \\ 
 & subgrid-scale models & to represent dynamics  & \\
\hline
Deep learning for accelerating & Deep learning /  & Autoencoders & Better interpretability  \\
the solution of the & reinforcement learning  & to learn manifolds & of deep learning for \\ 
Poisson equation & for more robust wall  & to represent dynamics & computational-fluid- \\
 & wall models  & & dynamics models \\
\hline
Domain reduction: deep learning & Deep learning / & Deep learning for  & More efficient  \\
for inflow  generation; Gaussian & symbolic regression & representing temporal  & operation of high- \\
processes for pressure-gradient & for more accurate and  & system dynamics & performance-computing\\
distributions & general RANS models & & facilities facilitated by ML\\
\hline
Deep learning for improved  & Deep learning / & Sparse identification  & ML-enabled more \\
accuracy of finite-difference / & symbolic regression & of nonlinear dynamics  & advanced computational  \\
finite-volume schemes & for interpretable & for learning efficient  & architectures \\
 & RANS models & and interpretable ROMs & \\
\hline
  \end{tabular}
  \caption{\label{table:summary}
  {\bf Summary of applications of ML to enhance CFD.} We show options to accelerate direct numerical simulations (DNSs), improvement of turbulence models for large-eddy simulations (LESs) and Reynolds-averaged Navier--Stokes (RANS) simulations, development of reduced-order models (ROMs) and possible future developments.
  }
  \end{center}
\end{table}

Physics-informed neural networks (PINNs) constitute another family of methods that are becoming widely adopted for scientific computing, more broadly. This technique, introduced by Raissi {\it et al.}~\cite{raissi2019physics}, uses deep learning to solve PDEs, exploiting the concept of automatic differentiation used in the back-propagation algorithm to calculate partial derivatives and form the equations, which are enforced through a loss function. Although the method is different from the ML approaches discussed above, in certain cases PINNs provide a promising alternative to traditional numerical methods for solving PDEs. This framework also shows promise for biomedical \textcolor{black}{simulations}, in particular after the recent work by Raissi {\it et al.}~\cite{raissi_et_al}, in which the concentration field of a tracer is used as an input to accurately predict the instantaneous velocity fields by minimizing the residual of the Navier--Stokes equations. \textcolor{black}{The growing usage of PINNs highlights the relevance of exploiting the knowledge we have about the governing equations when conducting flow simulations. In fact, PINNs are being used for turbulence modeling~\cite{pinns_etmm} (solving the RANS equations without the need of a model for the Reynolds stresses), for developing ROMs~\cite{kim_pinns}, for dealing with noisy data~\cite{pinns_exp} and for accelerating traditional solvers, for instance by means of solving the Poisson equation more efficiently~\cite{markidis}.}

There are also grand challenges in CFD that necessitate new methods in ML. One motivating \textcolor{black}{challenge} in CFD is to perform accurate coarse-resolution simulations in unforced three-dimensional wall-bounded turbulent flows. 
The production of turbulent kinetic energy (TKE) in these flows takes place close to the wall~\cite{kmm} (although at very high Reynolds numbers, outer-layer production also becomes relevant), and therefore using coarse meshes may significantly distort TKE production. In these flows of high technological importance, the TKE production sustains the turbulent fluctuations, and therefore a correction between coarse and fine resolutions may not be sufficient to obtain accurate results. Another challenge is the need for robust and interpretable models, a goal which is not easily attainable with deep-learning methods. There are however promising directions~\cite{cranmer_et_al,vinuesa_interp} to achieve interpretable deep-learning models, with important applications to CFD. A clear challenge of CFD is the large energy consumption associated with large-scale simulations. In this sense, ML can be an enabler for more efficient operation of the high-performance-computing (HPC) facilities~\cite{vinuesa_et_al_2020}, or even for future computational architectures, including quantum computers~\cite{koji_quantum}, as shown in Table~\ref{table:summary}.  

It is worth noticing that, while ML has a very high potential for CFD, there are also a number of caveats which may limit the applicability of ML to certain areas of CFD. Firstly, ML methods, such as deep learning, are often expensive to train and require large amounts of data. It is therefore important to identify areas where ML outperforms classical methods, which have been established for decades, and may be more accurate and efficient in certain cases. For instance, it is possible to develop interpretable ROMs with traditional methods such as POD and DMD, and while deep learning can provide some advantages~\cite{ae_modal}, the simpler machine-learning methods may be more efficient and straightforward. Finally, it is important to assess the information about the training data available to the user: certain flow properties ({\it e.g.} incompressibility, periodicity, etc.) should be embedded in the ML model to increase training efficiency and prediction accuracy. There is also a question of how the training data is generated, and whether the associated cost is taken into account when benchmarking. In this context, transfer learning is a promising area~\cite{guastoni2}.

Despite the caveats above, we believe that the trend of advancing CFD with ML will continue in the future.  This progress will continue to be driven by an increasing availability of high-quality data, high-performance computing, and a better understanding and facility with these emerging techniques. 
Improved adoption of reproducible research standards~\cite{barba2016hard, mesnard2017reproducible} is also a necessary step. 
Given the critical importance of data when developing ML modes, we advocate that the community continues to establish proper benchmark systems and best practices for open-source data and software in order to harness the full potential of ML to improve CFD.

\section*{Acknowledgements}
RV acknowledges the financial support from the Swedish Research Council (VR) and from the ERC Grant No. ``2021-CoG-101043998, DEEPCONTROL''. SLB acknowledges funding support from the Army Research Office (ARO W911NF-19-1-0045; program manager Dr. Matthew Munson).

\section*{Competing interests}
The authors declare no competing interests.

\section*{Author contributions}
\textcolor{black}{Both authors contributed equally to the ideation of the study and the writing process.}

 \begin{spacing}{.88}
 \setlength{\bibsep}{2.pt}

  \end{spacing}


\begin{thebibliography}{200}

\bibitem{godunov1959finite}
S.~Godunov and I.~Bohachevsky, ``Finite difference method for numerical
  computation of discontinuous solutions of the equations of fluid dynamics,''
  {\em Matemati{\v{c}}eskij sbornik}, vol.~47, no.~3, pp.~271--306, 1959.

\bibitem{eymard2000finite}
R.~Eymard, T.~Gallou{\"e}t, and R.~Herbin, ``Finite volume methods,'' {\em
  Handbook of Numerical Analysis}, vol.~7, pp.~713--1018, 2000.

\bibitem{zienkiewicz1977finite}
O.~C. Zienkiewicz, R.~L. Taylor, P.~Nithiarasu, and J.~Z. Zhu, {\em The finite
  element method}, vol.~3.
\newblock 1977.

\bibitem{canuto2012spectral}
C.~Canuto, M.~Y. Hussaini, A.~Quarteroni, and T.~A. Zang, {\em Spectral methods
  in fluid dynamics}.
\newblock Springer Science \& Business Media, 2012.

\bibitem{Brunton2019book}
S.~L. Brunton and J.~N. Kutz, {\em Data-Driven Science and Engineering: Machine
  Learning, Dynamical Systems, and Control}.
\newblock Cambridge University Press, 2019.

\bibitem{recht2019tour}
B.~Recht, ``A tour of reinforcement learning: The view from continuous
  control,'' {\em Annual Review of Control, Robotics, and Autonomous Systems},
  vol.~2, pp.~253--279, 2019.

\bibitem{vinuesa_et_al_2020}
R.~Vinuesa, H.~Azizpour, I.~Leite, M.~Balaam, V.~Dignum, S.~Domisch,
  A.~Fell{\"a}nder, S.~D. Langhans, M.~Tegmark, and F.~Fuso~Nerini, ``{The role
  of artificial intelligence in achieving the Sustainable Development Goals},''
  {\em Nature Communications}, vol.~11, p.~233, 2020.

\bibitem{Noe2020ARPC}
F.~No{\'e}, A.~Tkatchenko, K.-R. M{\"u}ller, and C.~Clementi, ``Machine
  learning for molecular simulation,'' {\em Annual review of physical
  chemistry}, vol.~71, pp.~361--390, 2020.

\bibitem{niederer2021scaling}
S.~A. Niederer, M.~S. Sacks, M.~Girolami, and K.~Willcox, ``Scaling digital
  twins from the artisanal to the industrial,'' {\em Nature Computational
  Science}, vol.~1, no.~5, pp.~313--320, 2021.

\bibitem{samuel}
A.~L. Samuel, ``Some studies in machine learning using the game of checkers,''
  {\em IBM Journal of Research and Development}, vol.~3, pp.~210--229, 1959.

\bibitem{Brenner2019prf}
M.~Brenner, J.~Eldredge, and J.~Freund, ``Perspective on machine learning for
  advancing fluid mechanics,'' {\em Physical Review Fluids}, vol.~4, no.~10,
  p.~100501, 2019.

\bibitem{Brunton2020arfm}
S.~L. Brunton, B.~R. Noack, and P.~Koumoutsakos, ``Machine learning for fluid
  mechanics,'' {\em Annual Review of Fluid Mechanics}, vol.~52, pp.~477--508,
  2020.

\bibitem{duraisamy_et_al}
K.~Duraisamy, G.~Iaccarino, and H.~Xiao, ``Turbulence modeling in the age of
  data,'' {\em Annual Review of Fluid Mechanics}, vol.~51, pp.~357--377, 2019.

\bibitem{ahmed2021closures}
S.~E. Ahmed, S.~Pawar, O.~San, A.~Rasheed, T.~Iliescu, and B.~R. Noack, ``On
  closures for reduced order models---a spectrum of first-principle to
  machine-learned avenues,'' {\em Physics of Fluids}, vol.~33, no.~9,
  p.~091301, 2021.

\bibitem{wang_wang}
B.~Wang and J.~Wang, ``Application of artificial intelligence in computational
  fluid dynamics,'' {\em Industrial \& Engineering Chemistry Research},
  vol.~60, pp.~2772--2790, 2021.

\bibitem{raissi2019physics}
M.~Raissi, P.~Perdikaris, and G.~E. Karniadakis, ``Physics-informed neural
  networks: A deep learning framework for solving forward and inverse problems
  involving nonlinear partial differential equations,'' {\em Journal of
  Computational Physics}, vol.~378, pp.~686--707, 2019.

\bibitem{karniadakis2021physics}
G.~E. Karniadakis, I.~G. Kevrekidis, L.~Lu, P.~Perdikaris, S.~Wang, and
  L.~Yang, ``Physics-informed machine learning,'' {\em Nature Reviews Physics},
  vol.~3, no.~6, pp.~422--440, 2021.

\bibitem{Noe2019science}
F.~No{\'e}, S.~Olsson, J.~K{\"o}hler, and H.~Wu, ``Boltzmann generators:
  Sampling equilibrium states of many-body systems with deep learning,'' {\em
  Science}, vol.~365, no.~6457, p.~eaaw1147, 2019.

\bibitem{vinuesa_ftac}
R.~Vinuesa, S.~M. Hosseini, A.~Hanifi, D.~S. Henningson, and P.~Schlatter,
  ``Pressure-gradient turbulent boundary layers developing around a wing
  section,'' {\em Flow Turbulence and Combustion}, vol.~99, pp.~613--641, 2017.

\bibitem{choi_moin}
H.~Choi and P.~Moin, ``{Grid-point requirements for large eddy simulation:
  Chapman’s estimates revisited},'' {\em Physics of Fluids}, vol.~24,
  p.~011702, 2012.

\bibitem{bar2019learning}
Y.~Bar-Sinai, S.~Hoyer, J.~Hickey, and M.~P. Brenner, ``Learning data-driven
  discretizations for partial differential equations,'' {\em Proceedings of the
  National Academy of Sciences}, vol.~116, no.~31, pp.~15344--15349, 2019.

\bibitem{stevens2020enhancement}
B.~Stevens and T.~Colonius, ``Enhancement of shock-capturing methods via
  machine learning,'' {\em Theoretical and Computational Fluid Dynamics},
  vol.~34, pp.~483--496, 2020.

\bibitem{jeon_kim}
J.~Jeon and S.~J. Kim, ``{FVM Network to reduce computational cost of CFD
  simulation},'' {\em Preprint arXiv:2105.03332}, 2021.

\bibitem{stevens2020finitenet}
B.~Stevens and T.~Colonius, ``Finitenet: A fully convolutional {LSTM} network
  architecture for time-dependent partial differential equations,'' {\em arXiv
  preprint arXiv:2002.03014}, 2020.

\bibitem{hoyer}
D.~Kochkov, J.~A. Smith, A.~Alieva, Q.~Wang, M.~P. Brenner, and S.~Hoyer,
  ``{Machine learning-accelerated computational fluid dynamics},'' {\em
  Proceedings of the National Academy of Sciences}, vol.~118, p.~e2101784118,
  2021.

\bibitem{chandler}
G.~J. Chandler and R.~R. Kerswell, ``{Invariant recurrent solutions embedded in
  a turbulent two-dimensional Kolmogorov flow},'' {\em Journal of Fluid
  Mechanics}, vol.~722, pp.~554--595, 2013.

\bibitem{bauer_et_al}
P.~Bauer, A.~Thorpe, and G.~Brunet, ``{The quiet revolution of numerical
  weather prediction},'' {\em Nature}, vol.~525, pp.~47--55, 2015.

\bibitem{freddy}
F.~Schenk, M.~V\"aliranta, F.~Muschitiello, L.~Tarasov, M.~Heikkil\"a,
  S.~Bj\"orck, J.~Brandefelt, A.~V. Johansson, J.~O. N\"aslund, and
  B.~Wohlfarth, ``{Warm summers during the Younger Dryas cold reversal},'' {\em
  Nature Communications}, vol.~9, p.~1634, 2018.

\bibitem{vinuesa_et_al_2018}
R.~Vinuesa, P.~S. Negi, M.~Atzori, A.~Hanifi, D.~S. Henningson, and
  P.~Schlatter, ``{Turbulent boundary layers around wing sections up to
  $Re_c=1,000,000$},'' {\em International Journal of Heat and Fluid Flow},
  vol.~72, pp.~86--99, 2018.

\bibitem{toras2018towards}
C.~Aloy~Tor{\'a}s, P.~Mimica, and M.~Mart{\'\i}nez-Sober, ``Towards detecting
  structures in computational astrophysics plasma simulations: using machine
  learning for shock front classification.,'' in {\em Artificial Intelligence
  Research and Development. Z. Falomir~{\it et al.} (Eds.)}, pp.~59--63, 2018.

\bibitem{li2020fourier}
Z.~Li, N.~Kovachki, K.~Azizzadenesheli, B.~Liu, K.~Bhattacharya, A.~Stuart, and
  A.~Anandkumar, ``Fourier neural operator for parametric partial differential
  equations,'' {\em arXiv preprint arXiv:2010.08895}, 2020.

\bibitem{li2020multipole}
Z.~Li, N.~Kovachki, K.~Azizzadenesheli, B.~Liu, K.~Bhattacharya, A.~Stuart, and
  A.~Anandkumar, ``Multipole graph neural operator for parametric partial
  differential equations,'' {\em arXiv preprint arXiv:2006.09535}, 2020.

\bibitem{li2020neural}
Z.~Li, N.~Kovachki, K.~Azizzadenesheli, B.~Liu, K.~Bhattacharya, A.~Stuart, and
  A.~Anandkumar, ``Neural operator: Graph kernel network for partial
  differential equations,'' {\em arXiv preprint arXiv:2003.03485}, 2020.

\bibitem{poisson2}
T.~Shan, W.~Tang, X.~Dang, M.~Li, F.~Yang, S.~Xu, and J.~Wu, ``{Study on a
  Poisson's equation solver based on deep learning technique},'' {\em 2017 IEEE
  Electrical Design of Advanced Packaging and Systems Symposium (EDAPS)},
  pp.~1--3, 2017.

\bibitem{poisson}
Z.~Zhang, L.~Zhang, Z.~Sun, N.~Erickson, R.~From, and J.~Fan, ``{Solving
  Poisson's equation using deep learning in particle simulation of PN
  junction},'' {\em 2019 Joint International Symposium on Electromagnetic
  Compatibility, Sapporo and Asia-Pacific International Symposium on
  Electromagnetic Compatibility (EMC Sapporo/APEMC)}, pp.~305--308, 2019.

\bibitem{bridson}
R.~Bridson, ``Fluid simulation,'' {\em A. K. Peters, Ltd., Natick, MA, USA},
  2008.

\bibitem{poisson3}
E.~Ajuria, A.~Alguacil, M.~Bauerheim, A.~Misdariis, B.~Cuenot, and E.~Benazera,
  ``{Towards a hybrid computational strategy based on deep learning for
  incompressible flows},'' {\em AIAA AVIATION Forum, June 15--19}, pp.~1--17,
  2020.

\bibitem{poisson_imperial}
A.~\"Ozbay, A.~Hamzehloo, S.~Laizet, P.~Tzirakis, G.~Rizos, and B.~Schuller,
  ``{Poisson CNN: Convolutional neural networks for the solution of the Poisson
  equation on a Cartesian mesh},'' {\em Data-Centric Engineering}, vol.~2,
  p.~E6, 2021.

\bibitem{Weymouth}
G.~D. Weymouth, ``Data-driven multi-grid solver for accelerated pressure
  projection,'' {\em arXiv preprint arXiv:2110.11029}, 2021.

\bibitem{fukami_inflow}
K.~Fukami, Y.~Nabae, K.~Kawai, and K.~Fukagata, ``{Synthetic turbulent inflow
  generator using machine learning},'' {\em Physical Review Fluids}, vol.~4,
  p.~064603, 2019.

\bibitem{morita_et_al}
Y.~Morita, S.~Rezaeiravesh, N.~Tabatabaei, R.~Vinuesa, K.~Fukagata, and
  P.~Schlatter, ``{Applying Bayesian optimization with Gaussian-process
  regression to computational fluid dynamics problems},'' {\em Journal of
  Computational Physics}, vol.~449, p.~110788, 2022.

\bibitem{boussinesq}
J.~V. Boussinesq, {\em {Th{\'e}orie analytique de la chaleur: mise en harmonie
  avec la thermodynamique et avec la th{\'e}orie m{\'e}canique de la
  lumi{\`e}re T. 2, Refroidissement et {\'e}chauffement par rayonnement
  conductibilit{\'e} des tiges, lames et masses cristallines courants de
  convection th{\'e}orie m{\'e}canique de la lumi{\`e}re}}.
\newblock Gauthier-Villars, 1923.

\bibitem{NASA}
J.~Slotnick, A.~Khodadoust, J.~Alonso, D.~Darmofal, W.~Gropp, E.~Lurie, and
  D.~Mavriplis, ``{CFD Vision 2030 study: a path to revolutionary computational
  aerosciences},'' {\em Technical Report NASA/CR–2014-218178}, 2014.

\bibitem{kutz2017deep}
J.~N. Kutz, ``Deep learning in fluid dynamics,'' {\em Journal of Fluid
  Mechanics}, vol.~814, pp.~1--4, 2017.

\bibitem{ling2016reynolds}
J.~Ling, A.~Kurzawski, and J.~Templeton, ``Reynolds averaged turbulence
  modelling using deep neural networks with embedded invariance,'' {\em Journal
  of Fluid Mechanics}, vol.~807, pp.~155--166, 2016.

\bibitem{craft_et_al}
T.~J. Craft, B.~E. Launder, and K.~Suga, ``Development and application of a
  cubic eddy-viscosity model of turbulence,'' {\em International Journal of
  Heat and Fluid Flow}, vol.~17, pp.~108--115, 1996.

\bibitem{hexagon}
O.~Marin, R.~Vinuesa, A.~V. Obabko, and P.~Schlatter, ``Characterization of the
  secondary flow in hexagonal ducts,'' {\em Physics of Fluids}, vol.~28,
  p.~125101, 2016.

\bibitem{spalart}
P.~R. Spalart, ``Strategies for turbulence modelling and simulations,'' {\em
  International Journal of Heat and Fluid Flow}, vol.~21, pp.~252--263, 2000.

\bibitem{wavy}
A.~Vidal, H.~M. Nagib, P.~Schlatter, and R.~Vinuesa, ``Secondary flow in
  spanwise-periodic in-phase sinusoidal channels,'' {\em Journal of Fluid
  Mechanics}, vol.~851, pp.~288--316, 2018.

\bibitem{wang2017physics}
J.~X. Wang, J.~L. Wu, and H.~Xiao, ``{Physics-informed machine learning
  approach for reconstructing Reynolds stress modeling discrepancies based on
  DNS data},'' {\em Physical Review Fluids}, vol.~2, p.~034603, 2017.

\bibitem{wu_et_al}
J.-L. Wu, H.~Xiao, and E.~Paterson, ``{Physics-informed machine learning
  approach for augmenting turbulence models: A comprehensive framework},'' {\em
  Physical Review Fluids}, vol.~3, p.~074602, 2018.

\bibitem{ml_rans}
C.~Jiang, R.~Vinuesa, R.~Chen, J.~Mi, S.~Laima, and H.~Li, ``An interpretable
  framework of data-driven turbulence modeling using deep neural networks,''
  {\em Physics of Fluids}, vol.~33, p.~055133, 2021.

\bibitem{rudin}
C.~Rudin, ``Stop explaining black box machine learning models for high stakes
  decisions and use interpretable models instead,'' {\em Nature Machine
  Intelligence}, vol.~1, pp.~206--215, 2019.

\bibitem{vinuesa_interp}
R.~Vinuesa and B.~Sirmacek, ``{Interpretable deep-learning models to help
  achieve the Sustainable Development Goals},'' {\em Nature Machine
  Intelligence}, vol.~3, p.~926, 2021.

\bibitem{cranmer_et_al}
M.~Cranmer, A.~Sanchez-Gonzalez, P.~Battaglia, R.~Xu, K.~Cranmer, D.~Spergel,
  and S.~Ho, ``{Discovering symbolic models from deep learning with inductive
  biases},'' {\em 34th Conference on Neural Information Processing Systems
  (NeurIPS 2020), Vancouver, Canada. Preprint arXiv:2006.11287}, 2020.

\bibitem{sandberg1}
J.~Weatheritt and R.~D. Sandberg, ``{A novel evolutionary algorithm applied to
  algebraic modifications of the RANS stress-strain relationship},'' {\em
  Journal of Computational Physics}, vol.~325, pp.~22--37, 2016.

\bibitem{gep}
J.~R. Koza, ``{Genetic Programming: On the Programming of Computers by Means of
  Natural Selection},'' {\em MIT Press}, 1992.

\bibitem{sandberg2}
J.~Weatheritt and R.~D. Sandberg, ``{The development of algebraic stress models
  using a novel evolutionary algorithm},'' {\em International Journal of Heat
  and Fluid Flow}, vol.~68, pp.~298--318, 2017.

\bibitem{Brunton2016pnas}
S.~L. Brunton, J.~L. Proctor, and J.~N. Kutz, ``Discovering governing equations
  from data by sparse identification of nonlinear dynamical systems,'' {\em
  Proceedings of the National Academy of Sciences}, vol.~113, no.~15,
  pp.~3932--3937, 2016.

\bibitem{beetham2020formulating}
S.~Beetham and J.~Capecelatro, ``Formulating turbulence closures using sparse
  regression with embedded form invariance,'' {\em Physical Review Fluids},
  vol.~5, no.~8, p.~084611, 2020.

\bibitem{schmelzer2020discovery}
M.~Schmelzer, R.~P. Dwight, and P.~Cinnella, ``Discovery of algebraic
  reynolds-stress models using sparse symbolic regression,'' {\em Flow,
  Turbulence and Combustion}, vol.~104, no.~2, pp.~579--603, 2020.

\bibitem{beetham2021sparse}
S.~Beetham, R.~O. Fox, and J.~Capecelatro, ``Sparse identification of
  multiphase turbulence closures for coupled fluid--particle flows,'' {\em
  Journal of Fluid Mechanics}, vol.~914, 2021.

\bibitem{rezaeiravesh}
S.~Rezaeiravesh, R.~Vinuesa, and P.~Schlatter, ``{On numerical uncertainties in
  scale-resolving simulations of canonical wall turbulence},'' {\em Computers
  and Fluids}, vol.~227, p.~105024, 2021.

\bibitem{iaccarino1}
M.~Emory, J.~Larsson, and G.~Iaccarino, ``{Modeling of structural uncertainties
  in Reynolds-averaged Navier--Stokes closures},'' {\em Physics of Fluids},
  vol.~25, p.~110822, 2013.

\bibitem{iaccarino2}
A.~A. Mishra and G.~Iaccarino, ``{Uncertainty estimation for Reynolds-averaged
  Navier--Stokes predictions of high-speed aircraft nozzle jets},'' {\em AIAA
  Journal}, vol.~55, pp.~3999--4004, 2017.

\bibitem{poroseva_et_al}
S.~Poroseva, F.~J.~D. Colmenares, and S.~Murman, ``{On the accuracy of RANS
  simulations with DNS data},'' {\em Physics of Fluids}, vol.~28, p.~115102,
  2016.

\bibitem{wu_et_al_2018}
J.~Wu, H.~Xiao, R.~Sun, and Q.~Wang, ``{Reynolds-averaged Navier--Stokes
  equations with explicit data-driven Reynolds stress closure can be
  ill-conditioned},'' {\em Journal of Fluid Mechanics}, vol.~869, pp.~553--586,
  2019.

\bibitem{accelerator}
O.~Obiols-Sales, A.~Vishnu, N.~Malaya, and A.~Chandramowlishwaran, ``{CFDNet: a
  deep learning-based accelerator for fluid simulations},'' {\em Preprint
  arXiv:2005.04485}, 2020.

\bibitem{sa}
P.~Spalart and S.~Allmaras, ``A one-equation turbulence model for aerodynamic
  flows,'' {\em 30th Aerospace Sciences Meeting and Exhibit, AIAA Paper
  1992-0439}, 1992.

\bibitem{openfoam}
H.~G. Weller, G.~Tabor, H.~Jasak, and C.~Fureby, ``A tensorial approach to
  computational continuum mechanics using object-oriented techniques,'' {\em
  Computers in Physics}, vol.~12, pp.~620--631, 1998.

\bibitem{multiphase}
F.~Gibou, D.~Hyde, and R.~Fedkiw, ``{Sharp interface approaches and deep
  learning techniques for multiphase flows},'' {\em Journal of Computational
  Physics}, vol.~380, pp.~442--463, 2019.

\bibitem{ma_et_al}
M.~Ma, J.~Lu, and G.~Tryggvasona, ``Using statistical learning to close
  two-fluid multiphase flow equations for a simple bubbly system,'' {\em
  Physics of Fluids}, vol.~27, p.~092101, 2015.

\bibitem{mi_et_al}
Y.~Mi, M.~Ishii, and L.~H. Tsoukalas, ``Flow regime identification methodology
  with neural networks and two-phase flow models,'' {\em Nuclear Engineering
  and Design}, vol.~204, pp.~87--100, 2001.

\bibitem{smagorinski}
J.~Smagorinsky, ``{General circulation experiments with the primitive
  equations: I. The basic experiment},'' {\em Monthly Weather Review}, vol.~91,
  pp.~99--164, 1963.

\bibitem{beck_et_al}
A.~D. Beck, D.~G. Flad, and C.-D. Munz, ``{Deep neural networks for data-driven
  {LES} closure models},'' {\em Journal of Computational Physics}, vol.~398,
  p.~108910, 2019.

\bibitem{lapeyre_et_al}
C.~J. Lapeyre, A.~Misdariis, N.~Cazard, D.~Veynante, and T.~Poinsot, ``Training
  convolutional neural networks to estimate turbulent sub-grid scale reaction
  rates,'' {\em Combustion and Flame}, vol.~203, p.~255, 2019.

\bibitem{maulik2019subgrid}
R.~Maulik, O.~San, A.~Rasheed, and P.~Vedula, ``Subgrid modelling for
  two-dimensional turbulence using neural networks,'' {\em Journal of Fluid
  Mechanics}, vol.~858, pp.~122--144, 2019.

\bibitem{kraichnan}
R.~H. Kraichnan, ``{Inertial ranges in two-dimensional turbulence},'' {\em
  Physics of Fluids}, vol.~10, pp.~1417--1423, 1967.

\bibitem{vollant}
A.~Vollant, G.~Balarac, and C.~Corre, ``{Subgrid-scale scalar flux modelling
  based on optimal estimation theory and machine-learning procedures},'' {\em
  Journal of Turbulence}, vol.~18, pp.~854--878, 2017.

\bibitem{gamahara}
M.~Gamahara and Y.~Hattori, ``{Searching for turbulence models by artificial
  neural network},'' {\em Physical Review Fluids}, vol.~2, p.~054604, 2017.

\bibitem{maulik}
R.~Maulik and O.~San, ``{A neural network approach for the blind deconvolution
  of turbulent flows},'' {\em Journal of Fluid Mechanics}, vol.~831,
  pp.~151--181, 2017.

\bibitem{sandberg3}
M.~Reissmann, J.~Hasslbergerb, R.~D. Sandberg, and M.~Klein, ``{Application of
  gene expression programming to a-posteriori LES modeling of a Taylor Green
  vortex},'' {\em Journal of Computational Physics}, vol.~424, p.~109859, 2021.

\bibitem{petros_natmi}
G.~Novati, H.~L. de~Laroussilhe, and P.~Koumoutsakos, ``{Automating turbulence
  modelling by multi-agent reinforcement learning},'' {\em Nature Machine
  Intelligence}, vol.~3, pp.~87--96, 2021.

\bibitem{hutchins}
N.~Hutchins, K.~Chauhan, I.~Marusic, J.~Monty, and J.~Klewicki, ``{Towards
  reconciling the large-scale structure of turbulent boundary layers in the
  atmosphere and laboratory},'' {\em Boundary-Layer Meteorology}, vol.~145,
  pp.~273--306, 2012.

\bibitem{log}
R.~E. Britter and S.~R. Hanna, ``{Flow and dispersion in urban areas},'' {\em
  Annual Review of Fluid Mechanics}, vol.~35, pp.~469--496, 2003.

\bibitem{giometto}
M.~G. Giometto, A.~Christen, C.~Meneveau, J.~Fang, M.~Krafczyk, and M.~B.
  Parlange, ``{Spatial characteristics of roughness sublayer mean flow and
  turbulence over a realistic urban surface},'' {\em Boundary-Layer
  Meteorology}, vol.~160, pp.~425--452, 2016.

\bibitem{bou_zeid}
E.~Bou-Zeid, C.~Meneveau, and M.~Parlange, ``{A scale-dependent Lagrangian
  dynamic model for large eddy simulation of complex turbulent flows},'' {\em
  Physics of Fluids}, vol.~17, p.~025105, 2005.

\bibitem{moeng}
C.~Moeng, ``{A large-eddy-simulation model for the study of planetary
  boundary-layer turbulence},'' {\em Journal of Atmospheric Sciences}, vol.~13,
  pp.~2052--2062, 1984.

\bibitem{mizuno_jimenez}
Y.~Mizuno and J.~Jim\'enez, ``{Wall turbulence without walls},'' {\em Journal
  of Fluid Mechanics}, vol.~723, pp.~429--455, 2013.

\bibitem{gmayoral}
M.~P. Encinar, R.~Garc\'ia-Mayoral, and J.~Jim\'enez, ``{Scaling of velocity
  fluctuations in off-wall boundary conditions for turbulent flows},'' {\em
  Journal of Physics: Conference Series}, vol.~506, p.~012002, 2014.

\bibitem{sasaki}
K.~Sasaki, R.~Vinuesa, A.~V.~G. Cavalieri, P.~Schlatter, and D.~S. Henningson,
  ``{Transfer functions for flow predictions in wall-bounded turbulence},''
  {\em Journal of Fluid Mechanics}, vol.~864, pp.~708--745, 2019.

\bibitem{ari_etmm}
G.~B. Arivazhagan, L.~Guastoni, A.~G\"uemes, A.~Ianiro, S.~Discetti,
  P.~Schlatter, H.~Azizpour, and R.~Vinuesa, ``{Predicting the near-wall region
  of turbulence through convolutional neural networks},'' {\em Proc. 13th
  ERCOFTAC Symp. on Engineering Turbulence Modelling and Measurements (ETMM13),
  Rhodes, Greece, September 16--17. Preprint arXiv:2107.07340}, 2021.

\bibitem{milano_koumoutsakos}
M.~Milano and P.~Koumoutsakos, ``Neural network modeling for near wall
  turbulent flow,'' {\em Journal of Computational Physics}, vol.~182,
  pp.~1--26, 2002.

\bibitem{moriya}
N.~Moriya, K.~Fukami, Y.~Nabae, M.~Morimoto, T.~Nakamura, and K.~Fukagata,
  ``{Inserting machine-learned virtual wall velocity for large-eddy simulation
  of turbulent channel flows},'' {\em Preprint arXiv:2106.09271}, 2021.

\bibitem{bae_koumoutsakos}
H.~J. Bae and P.~Koumoutsakos, ``{Scientific multi-agent reinforcement learning
  for wall-models of turbulent flows},'' {\em Nature Communications}, vol.~13,
  p.~1443, 2022.

\bibitem{Taira2017aiaa}
K.~Taira, S.~L. Brunton, S.~Dawson, C.~W. Rowley, T.~Colonius, B.~J. McKeon,
  O.~T. Schmidt, S.~Gordeyev, V.~Theofilis, and L.~S. Ukeiley, ``Modal analysis
  of fluid flows: An overview,'' {\em AIAA Journal}, vol.~55, no.~12,
  pp.~4013--4041, 2017.

\bibitem{Rowley2017arfm}
C.~W. Rowley and S.~T. Dawson, ``Model reduction for flow analysis and
  control,'' {\em Annual Review of Fluid Mechanics}, vol.~49, pp.~387--417,
  2017.

\bibitem{Taira2020aiaaj}
K.~Taira, M.~S. Hemati, S.~L. Brunton, Y.~Sun, K.~Duraisamy, S.~Bagheri,
  S.~Dawson, and C.-A. Yeh, ``Modal analysis of fluid flows: Applications and
  outlook,'' {\em AIAA Journal}, vol.~58, no.~3, pp.~998--1022, 2020.

\bibitem{lumley}
J.~L. Lumley, ``The structure of inhomogeneous turbulence,'' {\em Atmospheric
  turbulence and wave propagation, A. M. Yaglom and V. I. Tatarski (eds).
  Nauka, Moscow}, pp.~166--178, 1967.

\bibitem{Schmid2010jfm}
P.~J. Schmid, ``Dynamic mode decomposition of numerical and experimental
  data,'' {\em Journal of Fluid Mechanics}, vol.~656, pp.~5--28, Aug. 2010.

\bibitem{baldi1989neural}
P.~Baldi and K.~Hornik, ``Neural networks and principal component analysis:
  Learning from examples without local minima,'' {\em Neural Networks}, vol.~2,
  no.~1, pp.~53--58, 1989.

\bibitem{murata2020nonlinear}
T.~Murata, K.~Fukami, and K.~Fukagata, ``Nonlinear mode decomposition with
  convolutional neural networks for fluid dynamics,'' {\em Journal of Fluid
  Mechanics}, vol.~882, p.~A13, 2020.

\bibitem{ae_modal}
H.~Eivazi, S.~Le~Clainche, S.~Hoyas, and R.~Vinuesa, ``{Towards extraction of
  orthogonal and parsimonious non-linear modes from turbulent flows},'' {\em
  Expert Systems with Applications, To Appear. Preprint arXiv:2109.01514},
  2022.

\bibitem{Noack2003jfm}
B.~R. Noack, K.~Afanasiev, M.~Morzynski, G.~Tadmor, and F.~Thiele, ``A
  hierarchy of low-dimensional models for the transient and post-transient
  cylinder wake,'' {\em Journal of Fluid Mechanics}, vol.~497, pp.~335--363,
  2003.

\bibitem{lee2020model}
K.~Lee and K.~T. Carlberg, ``Model reduction of dynamical systems on nonlinear
  manifolds using deep convolutional autoencoders,'' {\em Journal of
  Computational Physics}, vol.~404, p.~108973, 2020.

\bibitem{Benner2015siamreview}
P.~Benner, S.~Gugercin, and K.~Willcox, ``A survey of projection-based model
  reduction methods for parametric dynamical systems,'' {\em SIAM Review},
  vol.~57, no.~4, pp.~483--531, 2015.

\bibitem{Carlberg2017jcp}
K.~Carlberg, M.~Barone, and H.~Antil, ``Galerkin v. least-squares
  petrov--galerkin projection in nonlinear model reduction,'' {\em Journal of
  Computational Physics}, vol.~330, pp.~693--734, 2017.

\bibitem{cenedese_et_al}
M.~Cenedese, J.~Ax\r{a}s, B.~B\"auerlein, K.~Avila, and G.~Haller,
  ``{Data-driven modeling and prediction of nonlinearizable dynamics via
  spectral submanifolds},'' {\em Nature Communications}, vol.~13, p.~872, 2022.

\bibitem{sole_conv}
M.~Lopez-Martin, S.~Le~Clainche, and B.~Carro, ``Model-free short-term fluid
  dynamics estimator with a deep {3D-convolutional} neural network,'' {\em
  Expert Systems With Applications}, vol.~177, p.~114924, 2021.

\bibitem{vlachas_et_al}
P.~R. Vlachas, W.~Byeon, Z.~Y. Wan, T.~P. Sapsis, and P.~Koumoutsakos,
  ``Data-driven forecasting of high-dimensional chaotic systems with long
  short-term memory networks,'' {\em Proceedings of the Royal Society A},
  vol.~474, p.~20170844, 2018.

\bibitem{srinivasan2019predictions}
P.~A. Srinivasan, L.~Guastoni, H.~Azizpour, P.~Schlatter, and R.~Vinuesa,
  ``Predictions of turbulent shear flows using deep neural networks,'' {\em
  Physical Review Fluids}, vol.~4, p.~054603, 2019.

\bibitem{sole_lstm}
R.~Abad\'ia-Heredia, M.~L\'opez-Mart\'in, B.~Carro, J.~I. Arribas, J.~M.
  P\'erez, and S.~Le~Clainche, ``A predictive hybrid reduced order model based
  on proper orthogonal decomposition combined with deep learning
  architectures,'' {\em Expert Systems With Applications}, vol.~187, p.~115910,
  2022.

\bibitem{pathak2018model}
J.~Pathak, B.~Hunt, M.~Girvan, Z.~Lu, and E.~Ott, ``Model-free prediction of
  large spatiotemporally chaotic systems from data: a reservoir computing
  approach,'' {\em Physical Review Letters}, vol.~120, no.~2, p.~024102, 2018.

\bibitem{Kaiser2014jfm}
E.~Kaiser, B.~R. Noack, L.~Cordier, A.~Spohn, M.~Segond, M.~Abel, G.~Daviller,
  J.~Osth, S.~Krajnovic, and R.~K. Niven, ``Cluster-based reduced-order
  modelling of a mixing layer,'' {\em J. Fluid Mech.}, vol.~754, pp.~365--414,
  2014.

\bibitem{peherstorfer2016data}
B.~Peherstorfer and K.~Willcox, ``Data-driven operator inference for
  nonintrusive projection-based model reduction,'' {\em Computer Methods in
  Applied Mechanics and Engineering}, vol.~306, pp.~196--215, 2016.

\bibitem{benner2020operator}
P.~Benner, P.~Goyal, B.~Kramer, B.~Peherstorfer, and K.~Willcox, ``Operator
  inference for non-intrusive model reduction of systems with non-polynomial
  nonlinear terms,'' {\em Computer Methods in Applied Mechanics and
  Engineering}, vol.~372, p.~113433, 2020.

\bibitem{qian2020lift}
E.~Qian, B.~Kramer, B.~Peherstorfer, and K.~Willcox, ``Lift \& learn:
  Physics-informed machine learning for large-scale nonlinear dynamical
  systems,'' {\em Physica D: Nonlinear Phenomena}, vol.~406, p.~132401, 2020.

\bibitem{Loiseau2017jfm}
J.-C. Loiseau and S.~L. Brunton, ``Constrained sparse {Galerkin} regression,''
  {\em Journal of Fluid Mechanics}, vol.~838, pp.~42--67, 2018.

\bibitem{loiseau2020data}
J.-C. Loiseau, ``Data-driven modeling of the chaotic thermal convection in an
  annular thermosyphon,'' {\em Theoretical and Computational Fluid Dynamics},
  vol.~34, no.~4, pp.~339--365, 2020.

\bibitem{guan2020sparse}
Y.~Guan, S.~L. Brunton, and I.~Novosselov, ``Sparse nonlinear models of chaotic
  electroconvection,'' {\em Royal Society Open Science}, vol.~8, no.~8,
  p.~202367, 2021.

\bibitem{Deng2020JFM}
N.~Deng, B.~R. Noack, M.~Morzynski, and L.~R. Pastur, ``Low-order model for
  successive bifurcations of the fluidic pinball,'' {\em Journal of Fluid
  Mechanics}, vol.~884, no.~A37, 2020.

\bibitem{deng2021galerkin}
N.~Deng, B.~R. Noack, M.~Morzy{\'n}ski, and L.~R. Pastur, ``Galerkin force
  model for transient and post-transient dynamics of the fluidic pinball,''
  {\em Journal of Fluid Mechanics}, vol.~918, 2021.

\bibitem{callaham2021empirical}
J.~L. Callaham, G.~Rigas, J.-C. Loiseau, and S.~L. Brunton, ``An empirical
  mean-field model of symmetry-breaking in a turbulent wake,'' {\em arXiv
  preprint arXiv:2105.13990}, 2021.

\bibitem{callaham2021role}
J.~L. Callaham, S.~L. Brunton, and J.-C. Loiseau, ``On the role of nonlinear
  correlations in reduced-order modeling,'' {\em arXiv preprint
  arXiv:2106.02409}, 2021.

\bibitem{Champion2019pnas}
K.~Champion, B.~Lusch, J.~N. Kutz, and S.~L. Brunton, ``Data-driven discovery
  of coordinates and governing equations,'' {\em Proceedings of the National
  Academy of Sciences}, vol.~116, no.~45, pp.~22445--22451, 2019.

\bibitem{Yeung2017arxiv}
E.~Yeung, S.~Kundu, and N.~Hodas, ``Learning deep neural network
  representations for koopman operators of nonlinear dynamical systems,'' {\em
  arXiv preprint arXiv:1708.06850}, 2017.

\bibitem{Takeishi2017nips}
N.~Takeishi, Y.~Kawahara, and T.~Yairi, ``Learning koopman invariant subspaces
  for dynamic mode decomposition,'' in {\em Advances in Neural Information
  Processing Systems}, pp.~1130--1140, 2017.

\bibitem{Lusch2018natcomm}
B.~Lusch, J.~N. Kutz, and S.~L. Brunton, ``Deep learning for universal linear
  embeddings of nonlinear dynamics,'' {\em Nature Communications}, vol.~9,
  no.~1, p.~4950, 2018.

\bibitem{Mardt2018natcomm}
A.~Mardt, L.~Pasquali, H.~Wu, and F.~No{\'e}, ``{VAMP}nets: Deep learning of
  molecular kinetics,'' {\em Nature Communications}, vol.~9, no.~5, 2018.

\bibitem{Otto2019siads}
S.~E. Otto and C.~W. Rowley, ``Linearly-recurrent autoencoder networks for
  learning dynamics,'' {\em SIAM Journal on Applied Dynamical Systems},
  vol.~18, no.~1, pp.~558--593, 2019.

\bibitem{wang2020incorporating}
R.~Wang, R.~Walters, and R.~Yu, ``Incorporating symmetry into deep dynamics
  models for improved generalization,'' {\em arXiv preprint arXiv:2002.03061},
  2020.

\bibitem{wang2020towards}
R.~Wang, K.~Kashinath, M.~Mustafa, A.~Albert, and R.~Yu, ``Towards
  physics-informed deep learning for turbulent flow prediction,'' in {\em
  Proceedings of the 26th ACM SIGKDD International Conference on Knowledge
  Discovery \& Data Mining}, pp.~1457--1466, 2020.

\bibitem{frezat2021physical}
H.~Frezat, G.~Balarac, J.~Le~Sommer, R.~Fablet, and R.~Lguensat, ``Physical
  invariance in neural networks for subgrid-scale scalar flux modeling,'' {\em
  Physical Review Fluids}, vol.~6, no.~2, p.~024607, 2021.

\bibitem{erichson2019physics}
N.~B. Erichson, M.~Muehlebach, and M.~W. Mahoney, ``Physics-informed
  autoencoders for lyapunov-stable fluid flow prediction,'' {\em arXiv preprint
  arXiv:1905.10866}, 2019.

\bibitem{kaptanoglu2021promoting}
A.~A. Kaptanoglu, J.~L. Callaham, C.~J. Hansen, A.~Aravkin, and S.~L. Brunton,
  ``Promoting global stability in data-driven models of quadratic nonlinear
  dynamics,'' {\em Physical Review Fluids}, vol.~6, no.~094401, 2021.

\bibitem{drl_control}
R.~Vinuesa, O.~Lehmkuhl, A.~Lozano-Dur\'an, and J.~Rabault, ``Flow control in
  wings and discovery of novel approaches via deep reinforcement learning,''
  {\em Fluids}, vol.~865, pp.~281--302, 2019.

\bibitem{guastoni2}
L.~Guastoni, A.~G\"uemes, A.~Ianiro, S.~Discetti, P.~Schlatter, H.~Azizpour,
  and R.~Vinuesa, ``Convolutional-network models to predict wall-bounded
  turbulence from wall quantities,'' {\em Journal of Fluid Mechanics},
  vol.~928, p.~A27, 2021.

\bibitem{kim910unsupervised}
H.~Kim, J.~Kim, S.~Won, and C.~Lee, ``Unsupervised deep learning for
  super-resolution reconstruction of turbulence,'' {\em Journal of Fluid
  Mechanics}, vol.~910, p.~A29, 2021.

\bibitem{fukami2019super}
K.~Fukami, K.~Fukagata, and K.~Taira, ``Super-resolution reconstruction of
  turbulent flows with machine learning,'' {\em Journal of Fluid Mechanics},
  vol.~870, pp.~106--120, 2019.

\bibitem{guemes_gans}
A.~G\"uemes, S.~Discetti, A.~Ianiro, B.~Sirmacek, H.~Azizpour, and R.~Vinuesa,
  ``From coarse wall measurements to turbulent velocity fields through deep
  learning,'' {\em Physics of Fluids}, vol.~33, p.~075121, 2021.

\bibitem{fukami_autoencoders}
K.~Fukami, T.~Nakamura, and K.~Fukagata, ``Convolutional neural network based
  hierarchical autoencoder for nonlinear mode decomposition of fluid field
  data,'' {\em Physics of Fluids}, vol.~32, p.~095110, 2020.

\bibitem{raissi_et_al}
M.~Raissi, A.~Yazdani, and G.~E. Karniadakis, ``{Hidden fluid mechanics:
  Learning velocity and pressure fields from flow visualizations},'' {\em
  Science}, vol.~367, pp.~1026--1030, 2020.

\bibitem{pinns_etmm}
H.~Eivazi, M.~Tahani, P.~Schlatter, and R.~Vinuesa, ``{Physics-informed neural
  networks for solving Reynolds-averaged Navier--Stokes equations},'' {\em
  Proc. 13th ERCOFTAC Symp. on Engineering Turbulence Modelling and
  Measurements (ETMM13), Rhodes, Greece, September 16--17. Preprint
  arXiv:2107.10711}, 2021.

\bibitem{kim_pinns}
Y.~Kim, Y.~Choi, D.~Widemann, and T.~Zohdi, ``A fast and accurate
  physics-informed neural network reduced order model with shallow masked
  autoencoder,'' {\em Journal of Computational Physics}, p.~110841, 2021.

\bibitem{pinns_exp}
H.~Eivazi and R.~Vinuesa, ``Physics-informed deep-learning applications to
  experimental fluid mechanics,'' {\em Preprint arXiv:2203.15402}, 2022.

\bibitem{markidis}
S.~Markidis, ``The old and the new: can physics-informed deep-learning replace
  traditional linear solvers?,'' {\em Preprint arXiv:2103.09655v2}, 2021.

\bibitem{kmm}
J.~Kim, P.~Moin, and R.~Moser, ``{Turbulence statistics in fully developed
  channel flow at low Reynolds number},'' {\em Journal of Fluid Mechanics},
  vol.~177, pp.~133--166, 1987.

\bibitem{koji_quantum}
K.~Fukagata, ``{Towards quantum computing of turbulence},'' {\em Nature
  Computational Science}, vol.~2, pp.~68--69, 2022.

\bibitem{barba2016hard}
L.~A. Barba, ``The hard road to reproducibility,'' {\em Science}, vol.~354,
  no.~6308, pp.~142--142, 2016.

\bibitem{mesnard2017reproducible}
O.~Mesnard and L.~A. Barba, ``Reproducible and replicable computational fluid
  dynamics: it's harder than you think,'' {\em Computing in Science \&
  Engineering}, vol.~19, no.~4, pp.~44--55, 2017.

\end{thebibliography}
\end{document}